\documentclass[runningheads]{cl2emult}

\usepackage{makeidx}  
\usepackage{graphicx} 
\usepackage{subeqnar} 
\usepackage{multicol} 
\usepackage{cropmark} 
\usepackage{lnp}      
\makeindex            

\begin{document}
\title*{A Deep 12$\mu$m Survey with ISO}
\toctitle{A Deep 12$\mu$m Survey with ISO}
%
%
\titlerunning{ISO 12$\mu$m Survey}
%
\author{D.L. Clements\inst{1,2}}
\authorrunning{D.L. Clements}
%
%
\institute{Dept. Physics and Astronomy, Cardiff University, PO Box 913,
Cardiff, CF2 3YB, UK
\and Institut d'Astrophysique Spatiale, Universit\'{e} de Paris XI, 
B\^{a}timent 121,
F-91405 Orsay Cedex, France}

\maketitle              

\begin{abstract}
I present results from a deep 12$\mu$m extragalactic survey conducted
with the ISOCAM instrument. The survey covers about 0.1 sq. deg. in
four fields and reaches a 5$\sigma$ flux limit of $\sim$500$\mu$Jy. 50
sources are identified to this flux limit. Of these, 37 are classified
as galaxies on the basis of optical/mid-IR colours using
identifications from the USNO-A photographic survey. Number counts for
these objects exceed those predicted for no-evolution models in simple
models.  However, these conclusions are somewhat dependent on the
assumed K-corrections. For this reason, and to better determine the
nature of the evolution of this population, followup observations are
required to determine redshifts, broadband optical-IR colours, and
optical morphologies. The first results from these followups are
presented.  Images and optical/IR photometry for one of the four
fields is discussed, and I also present the first results from optical
spectroscopy. The highest redshift for the sample so far is z=1.2 for
a broad-line object.
\end{abstract}

\section{Introduction}

Our view of the universe was fundamentally changed when the IRAS
satellite revealed the importance of the mid- and far-IR parts of the
spectrum in the light output of normal galaxies. About 1/3 of the
luminosity of normal spirals is emitted in the mid- to far-IR (Soifer
\& Neugebauer, 1991). Perhaps more surprising was the discovery of a
class of objects with very high luminosities, $> 10^{12}L_{\odot}$,
which is almost entirely emitted in the mid- to far-IR (eg. Joseph \&
Wright, 1985; Sanders \& Mirabel, 1996, and references therein). Such
objects are known as Ultraluminous Infrared Galaxies, and have been
the targets of intense study in the local universe (eg. Genzel, these
proceedings). The issue of evolution in the ULIRG and broader IR
luminous populations has always been of interest. However, it was not
possible to observationally tackle this issue until recently since our
main database on such objects, the IRAS catalogues, were dominated by
relatively low redshift sources. Two instruments have now changed this
situation --- ISO, and SCUBA. SCUBA provides excellent
possibilities for deep surveys in the submm, allowing us to look for
high redshift counterparts to local ULIRGs (eg. Eales et al., 1999 and
references therein). ISO allows deeper surveys in the mid- and far-IR,
boosting our capabilities relative to IRAS by up to 1000 times. The
whole issue of far-IR evolution has also been made more urgent by the
discovery of a cosmological infrared background (eg. Hauser, Lagache,
Puget contributions in this volume).

There are numerous mid- and far-IR surveys underway, many of which are
discussed in this volume. The two instruments on ISO that permit such
surveys are ISOCAM, operating in the mid-IR (Cesarsky et al. 1996),
and ISOPHOT, operating in the far-IR (Lemke at al., 1996). There are
arguments both for and against the use of both instruments and surveys
have been undertaken with both (see eg. Dole, Cesarsky, Oliver
contributions in this volume). We here discuss the results from a
survey conducted with ISOCAM using the LW10, 12$\mu$m filter.

\section{Why a 12$\mu$m Survey?}

By far the bulk of the surveys conducted with ISOCAM have been made at
7 or 15$\mu$m. Such surveys cover larger areas of the sky (eg. ELAIS,
Oliver, this volume) and go deeper (eg. ULTRADEEP, Cesarsky, this
volume) than the present work. One might then ask why we should bother
with a 12$\mu$m survey. The answer to this lies in two
directions. Firstly, the LW10 filter broadly matches the IRAS 12$\mu$m
passband. Our current survey can thus act as a bridge between the
shallower all-sky surveys from IRAS with the deeper, smaller area
surveys at 7 and 15$\mu$m conducted with ISO. Secondly, the mid-IR
emission of galaxies contains a complex mix of different emission
mechanisms.  These include the unidentified infrared bands (eg. at 7.7
and 11.3$\mu$m), various emission line species (eg. NeII at 12.7$\mu$m),
absorption lines (principally the silicate feature at 9.7$\mu$m), and
contributions from very hot dust grains at $\sim$200K (see eg. Aussel
et al., 1999). When this complex spectral energy distribution (SED) is
combined with the effects of redshift and specific observational
filters, there is the potential for substantial colour(K)-corrections
to any survey.  The broader the observational filter, the less subject
to K-corrections the resulting survey becomes. Since we do not know,
and, until SIRTF observations arrive, will not be able to determine,
the underlying mid-IR SED of any given galaxy in a mid-IR survey, this
can lead to significant uncertainties in scientific conclusions. This
is especially true when the redshifts of the sources are unknown. Xu
et al. (1998) have shown that both the average K-corrections for a
12$\mu$m ISO survey and the uncertainties in those corrections for any
individual object are significantly smaller than for 15$\mu$m
observations. The present observations will thus be easier to
interpret thanks to both the smaller K-corrections, and the copious
amounts of IRAS 12$\mu$m data available.

\section{The ISO Deep 12$\mu$m Survey: A Happy Accident}

The data for the 12$\mu$m survey were not originally acquired for this
purpose.  They were originally obtained for a search for comet trails
associated with comet 7P/Pons-Winnecke, and were part of a larger
project (see Davies et al., 1997). The original aim was to image four
fields with ISOCAM raster maps in regions 1$^o$ ahead and 0.5$^o$,
1$^o$ and 2$^o$ behind the nucleus of the comet. Each raster was 11x7
pointings, with 6'' pixel-field-of-view, and 60'' by 48''
spacing. Each position on the sky was thus visited 12 times, providing
good redundancy to the data. Total integration time per position is
about 300s. Unfortunately, the observations were scheduled one day
later than assumed in the ephemeris calculations for the comet,
ensuring that the comet trails will lie at or beyond the bottom of
each image. No trace of the trails is in fact seen in the reduced
data, so we are left with four fields at high galactic latitude
(typically $b= - 53$) with deep mid-IR imaging --- ideal for a
cosmological survey. The data were reduced using the IAS
dual-beamswitch method (Desert et al. 1999). Further details of the
data reduction are discussed in Clements et al. (1999).

\section{Survey Results}

\begin{figure}
\centering
\includegraphics[width=.7\textwidth]{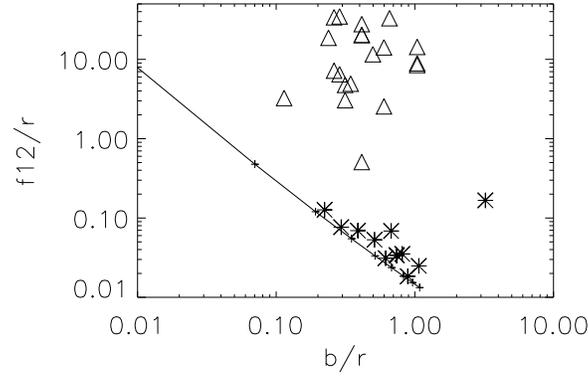}
\caption[]{Optical-IR Colour-Colour Diagram}
Triangles are 12$\mu$m galaxies, stars are objects identified as stars
on the basis of their optical-IR colours. The line shows where a pure
Rayleigh-Jeans spectrum object would lie. The star on the far right of the
plot is a merged pair of stars in the 12$\mu$m data.
\label{eps1}
\end{figure}

\begin{figure}
\centering
\includegraphics[width=.5\textwidth,angle=90]{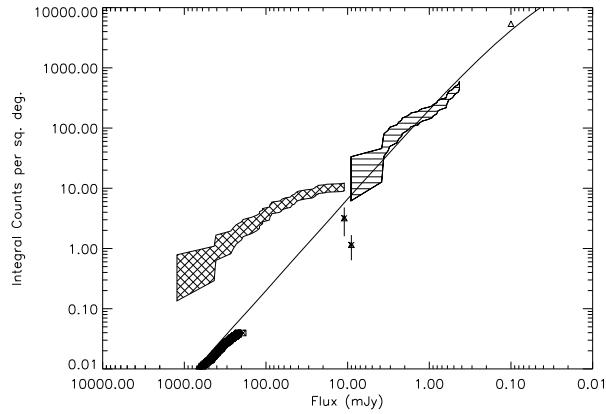}
\caption[]{12$\mu$m Integrated Number Counts}
Single hatched region are the number counts from the present data,
double hatched are counts from the deepest IRAS 12$\mu$m survey
(Hacking \& Houck, 1987) which is dominated by stars, dark line to the
bottom are counts from the IRAS galaxy survey (Rush et
al. 1993). Triangles in the middle of the diagram are galaxies from
Hacking \& Houck survey (lower) and Gregorich et al. (1995). Triangle
to upper right are the 15$\mu$ counts from the ISOHDF (Oliver et al.,
1997). Solid line is a strong evolution model from Clements et al. (1999).
\label{eps2}
\end{figure}

A total of 148 verified sources are detected to 3$\sigma$ in the
survey, down to a flux limit of 300$\mu$Jy. For number count purposes
we restrict ourselves to those sources detected at $>5\sigma$ for a
number of reasons. Firstly a number of uncertainties remain in the
identification of the weakest sources, and secondly the problems of
Malmquist bias are most easily controlled in samples detected at $>
5\sigma$.  50 objects are detected at $>5 \sigma$. We obtain optical
identifications for these objects from the USNO-A catalogue, which
includes B and R band magnitudes. We remove stars from this catalog on
the basis of optical-IR colours (see Fig. \ref{eps1}), and then plot the
integrated number counts for the galaxies alone (Fig. \ref{eps2}).

The number counts from this plot can be compared to evolving and
non-evolving models. We find that simple non-evolving models are
excluded by this data at the 3.5$\sigma$ level. However, uncertainties
in the K-corrections mean that this conclusion is not final. For that,
we need further data including redshifts.

\section{Optical/IR Colours and Imaging}

\begin{figure}
\centering
\includegraphics[width=.7\textwidth,angle=90]{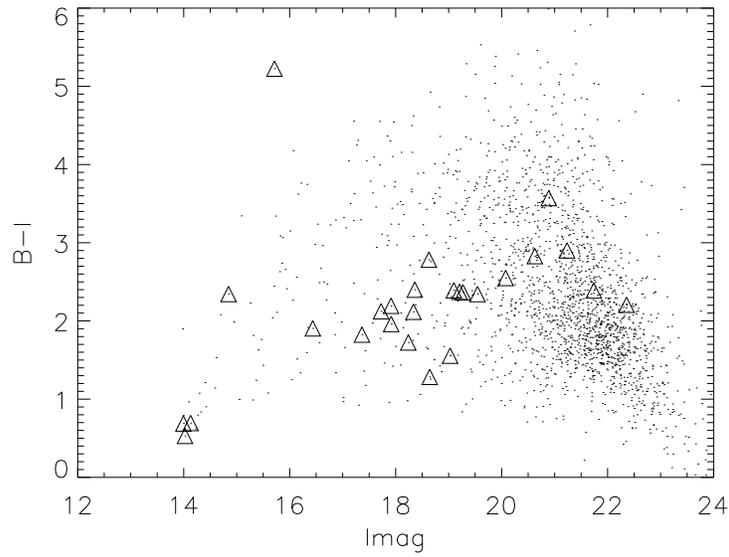}
\caption[]{B-I Colour Magnitude Diagram}
Triangles are 12$\mu$m sources, the brightest of which are stars. Dots
are all the optically selected objects in the field. Note that the
12$\mu$m sources, with one exception, are not unusually red,
indicating that the optical emission of these objects is not heavily
obscured by dust. The one unusually red object turns out to be an M
star.
\label{eps3}
\end{figure}

As part of the UKIRT mini-survey and the INT Wide Field Survey
programme, much of the first of the four 12$\mu$m fields was imaged at UBVRIJ
and K. This multicolour imaging allows us to examine both the
morphologies and optical/IR SEDs of these objects. We include in this
study not only those objects detected at $>5\sigma$ but also those
detected at $>3\sigma$ with believable optical counterparts. The issue
of any blank field 12$\mu$m sources, ie. those detected at $>3\sigma$
but with no optical counterpart, will be discussed elsewhere. We find
that the optical/near-IR colours of the 12$\mu$m sources are largely
unexceptional (see Fig. \ref{eps3}). Only one source has an unusually
large B-I colour (later spectroscopy shows this object to be an M-type
star). The rest of the sources largely have colours typical of the
bulk of the population seen in this field. One must then ask why these
objects have become luminous at 12$\mu$m. The optical images perhaps
provide a clue to this (Fig 4). Inspection of these reveals
that 77\% of the 12$\mu$m galaxies have companions or disturbed
morphologies. This compares with a rate of $\sim$ 10\% for low
luminosity IRAS galaxies (Lawrence et al. 1987?), and suggests that
interactions or mergers may be involved in triggering the 12$\mu$m
activity of these sources.

\begin{figure}
\centering
\includegraphics[width=\textwidth]{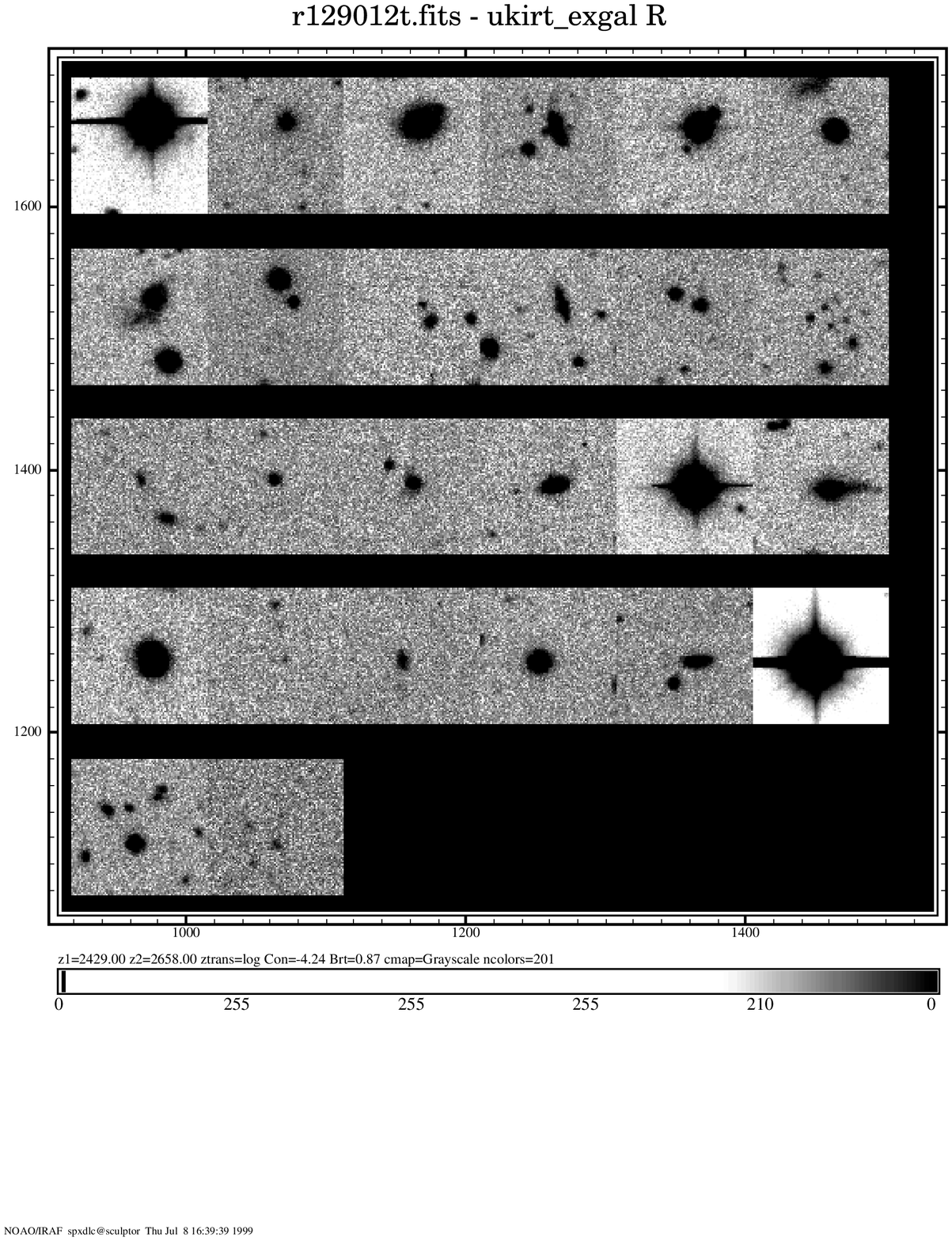}
\caption[]{R band images for the 26 12$\mu$m sources found in Field 1}
Images are 37'' across with 0.37'' pixels.
\label{eps4}
\end{figure}

\section{Optical Spectroscopy}

\begin{figure}
\centering
\includegraphics[width=\textwidth]{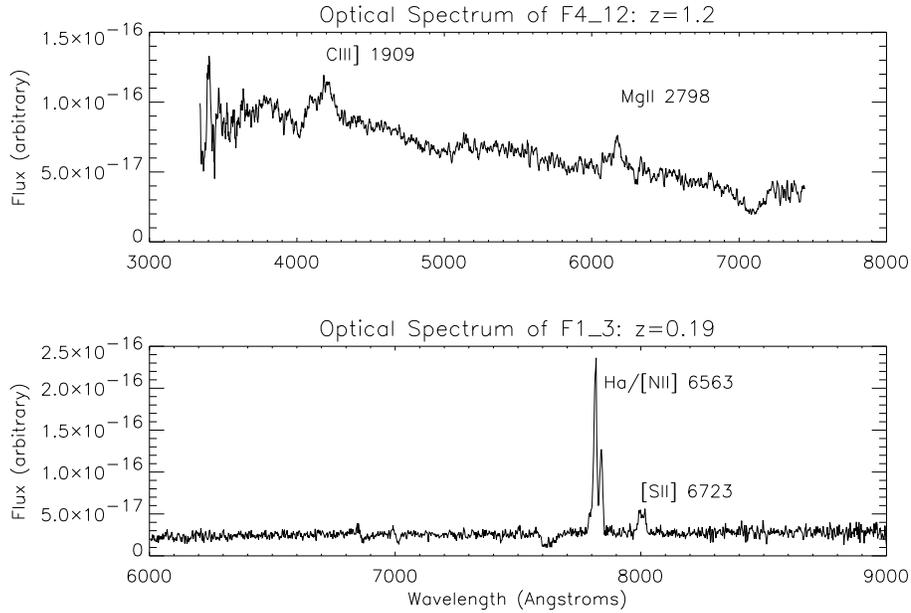}
\caption[]{Example spectra for two sources in the 12$\mu$m survey}
\label{eps5}
\end{figure}

Of the 12$\mu$m sources identified in this survey, only one had a
previously known redshift (z=0.11 (Clements et al., 1996)). We must
obtain redshifts for the whole sample, or complete subsamples, to
properly test evolutionary models. Optical spectroscopy for these
sources is thus a high priority. We have so far had two runs at the
ESO 3.6m telescope using EFOSC-2 for this part of the programme. The
first of these runs suffered from poor weather conditions and
instrument failures. In contrast, the most recent of these runs,
shortly after the Ringberg meeting of which these are the proceedings,
went very well. We now have secure redshifts for at least 25 objects,
with redshifts ranging from 0.037 to 1.2. We have also found that one
or two of the objects previously identified as galaxies turn out to be
M-stars. Their SEDs appear to be galaxy like in the optical-mid-IR
colour-colour diagram because of the large molecular absorption bands
reducing their optical emission below Rayleigh-Jeans levels. As well
as redshift determination, we will use these spectra to classify the
nature of the ionising source in the objects. It already appears that
the majority of these sources are powered by star formation, since
they have HII region-like spectra. Several, though, show signs of an
AGN contribution, including our highest redshift object which has a
broad line spectrum (Fig. \ref{eps5}).

\section{Conclusions}

I have presented results from a deep 12$\mu$m survey with ISO. The
survey data themselves suggest that strong evolution is taking place
in this population, but additional followup observations are needed
to confirm this. The followup programme is now well underway, with
optical/IR imaging for one of the four survey areas in hand, and
optical spectra complete for the brighter objects in the sample. The
next stage will be to construct luminosity functions for the
spectroscopic subsample and to compare this with IRAS data on nearby
objects. With the sad demise of the WIRE spacecraft, the present work
is likely to be the deepest 12$\mu$m survey available until the NGST
era. As such it represents an important link between the IRAS surveys
and the deeper, smaller area ISO surveys at 7 and 15$\mu$m. It is thus an
important resource for the future of mid-IR cosmology.
\\~\\
{\bf Acknowledgements} It is a pleasure to thank my collaborators on
this project: Xavier Desert, Alberto Franceschini, Bill Reach, Amanda
Baker, John Davies and Catherine Cesarsky. I am also very grateful to
Steve Warren, Scott Croom and the UKIRT Minisurvey team for providing
the optical/IR imaging data. Thanks also to ESA and ESO for the use of
their facilities. This work was supported by EU TMR network and
PPARC postdoctoral posts.

\clearpage
\addcontentsline{toc}{section}{Index}
\flushbottom
\printindex

\end{document}